\documentclass[11pt]{article}
\usepackage{epsf}

\title{Application of Polynomial Algorithms to a Random Elastic Medium} 
 
\author{Chen Zeng and P.L. Leath\\
\small Department of Physics, 
Rutgers University, Piscataway, NJ 08854, USA} 
\date{}  
\begin{document}
\maketitle  

\begin{abstract}  
A randomly pinned elastic medium in two dimensions is 
modeled by a disordered fully-packed loop model. The energetics of   
disorder-induced dislocations is studied using exact
and polynomial algorithms from combinatorial optimization. 
Dislocations are found to become unbound at large scale, and
the elastic phase is thus unstable giving evidence 
for the absence of a Bragg glass in two dimensions. 
\end{abstract}


Randomly pinned elastic media are used to model various condensed-matter
systems with quenched disorder including the vortex phase of dirty type-II 
superconductors\cite{Blatter}. Much analytical progress on these systems has been 
made within the elastic approximation where dislocations are excluded 
by fiat. The intriguing possibility of spontaneous formation of disorder-induced
dislocations (pairs and loops respectively in two and three dimensions) 
at large scale, however, remains a challenging question\cite{BG}. 
To address this issue at zero temperature 
requires a detailed understanding of the energetics 
of dislocations in terms of their elastic-energy cost and  
disorder-energy gain in the ground state{BG}. 

In recent years, we have witnessed a fruitful exploration of novel algorithms 
in for complex disordered systems whose ground state itself 
is dominated by random disorder. One class of these efficient algorithms 
is based on network flow optimization. It includes the min-cost-flow, 
max-flow and matching algorithms which compute the exact ground state 
in time that grows only polynomialy in the system size, an attibution 
of great practical importance. Some recent applications include studies 
of the roughness and topography of random manifolds and $2d$ random
elastic media by max-flow\cite{max1}, matching\cite{matching1} 
and min-cost-flow algorithms\cite{min1},    
the sensitivity exponents of the random-field-Ising model by max-flow 
algorithms\cite{max2}, the domain-wall energy in the gauge glass by  
min-cost-flow algorithms\cite{min2}, the droplet excitations in disordered 
systems by matching algorithms\cite{matching2}, and the critical exponents of 
$2d$ generic rigidity percolation by matching algorithms\cite{percolation}. 


In this article we briefly review our recent work\cite{dislocation1} 
on dislocations in $2d$ randomly pinned elastic media and show how the 
energetics of dislocation pairs can be studied numerically 
by applying these polynomial algorithms to a $2d$ lattice model. 
The essential ingredients required of such a $2d$ discrete model would be:      
(1) its large-scale fluctuations are described by an elastic Hamiltonian 
with a quenched random potential that reflects the periodicity intrinsic
to any elastic medium; (2) dislocations can be ``conveniently'' generated; 
and (3) its ground-state energies with and without dislocations are
amenable to exact numerical computations by these polynomial algorithms. 
 

\paragraph{Models} Fortunately, recent works of Henley, Kondev and their 
co-workers provided us with a large class of just such models whose degrees 
of freedom are described in terms of colors, tilings (dimer) and loops\cite{height},  
precisely the natural language for considering network flow optimizations.
More importantly, these models all permit a solid-on-solid (SOS) representation 
whose large-scale height fluctuations are governed by a few elastic constants  
and a locking potential that is periodic in heights. As an illustration, 
we consider here a fully-packed loop (FPL) model defined on a honeycomb lattice. 
All configurations of occupied bonds which form closed loops and cover every site 
exactly once are allowed, as in the example of Fig.1(a). The corresponding SOS surface 
is a (111)-interface of a simple cubic lattice constructed as follows.
Define integer heights at the centers of the hexagons of this honeycomb
lattice then orient 
all bonds of the resulting triangular lattice connecting the centers
such that elementary triangles
pointing upward are circled clockwise; assign $+1$ to the difference of
neighboring heights along the oriented bonds if a loop is crossed and $-2$
otherwise. This yields single-valued heights up to an overall constant. 

We introcude quenched disorder via random bond weights on the honeycomb lattice,
chosen independently and uniformly from integers in the interval [$-w$, $w$] with 
$w=500$. The total energy is the sum of the bond weights along all loops and 
strings. The FPL model is shown to be 
equivalent to an array of fluxlines confined in a plane\cite{matching1} 
with the heights corresponding to the displacement fields of the fluxlines.  
The SOS surface described above can be viewed as an elastic surface 
embedded in a $3d$ random potential that is periodic in heights modulo 3 
since the smallest ``step'' of the surface is three\cite{height}.  
The coarse-grained effective Hamiltonian becomes\cite{dislocation1,height,private}  
\begin{equation}
H=\int d{\bf r}
\left[
\frac{K}{2} \left( \nabla h({\bf r}) \right)^2
- u\cos\left(\frac{2\pi}{3} h({\bf r}) - \gamma({\bf r})\right)
\right]
\;\; , 
\label{eq1}
\end{equation}
where the random bond weights enter as random phase $\gamma({\bf r})$. 
Note also both $K$ and $u$ depend on the disorder strength $w$ since it is 
the only energy scale in the problem. This is the well studied model 
for charge-density waves (CDW)\cite{CDW}. 

Dislocations are added to the FPL model by ``violating'' the fully-packed constraint. 
One dislocation pair is an open string in an otherwise fully-packed 
system as shown in Fig.1(b). The height change along any path encircling 
one end of the string is the Burgers charge $\pm 3$ of a dislocation so
that the heights become multi-valued. Note that the configurations with 
and without a dislocation pair only differ along a domain ``wall'', 
as shown in Fig.1(c). Dislocations with higher Burgers charges $\pm 6$ 
can also be created by introducing holes instead of strings.   

\begin{figure}
  \begin{center}
    \leavevmode
    \epsfxsize=8cm
    \epsffile{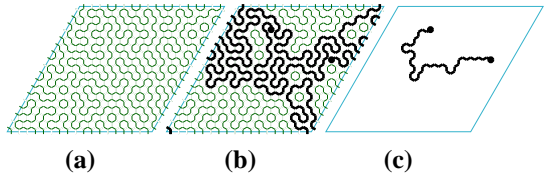}
  \end{center}
\caption{
The FPL model with periodic boundary conditions. The ground states 
with and without a dislocation pair for one realization of random bond 
weights are displayed in (b) and (a) respectively. 
The dislocations (solid dots) in (b) are connected by an open string
(thick line) among the loops. The relevant physical object is, however,
the domain wall which is induced by the dislocations as shown in (c). This
domain wall represents the line of all bond {\it differences} between the
ground states (a) and (b).
}
\label{fig_1}
\end{figure}


\paragraph{Algorithms}
It turns out that the ground states of the disordered FPL model can 
be obtained via polynomial algorithms. A general description of 
such an optimization problem is given by the so-called linear programming 
which is to identify a set of variables minimizing a linear objective  
function subject to a set of linear constraints. Most physical problems 
are restricted to integer-valued variables. Such an integer optimization problem
in general is nondeterministic polynomial (NP) which implies that 
polynomial and exact algorithms are unlikely to be found.   
However, for a special class of problems where the linear constraints other than 
the upper and lower bounds on the variables can be interpreted
as ``flow conservation'' at the nodes of a graph while the variables 
are identified with the flows on the edges of the graph, the optimization 
problem is polynomial. Recent applications mentioned above fall into 
this class. Since most textbooks\cite{OP} contain details on 
the proof of this result and existing C++ codes for these polynomial algorithms 
can be found in the LEDA library\cite{LEDA}, we shall only discuss how 
to transform the search for the ground states into an integer 
min-cost-flow problem on a suitably designed graph. 
The min-cost-flow problem is to find the flow 
pattern of minimum total cost for sending a specific amount of 
flow from a given node $s$ to another given node $t$ in a 
graph $G$ in which the flow $x$ on every edge has an upper bound $u_b$ and
a lower bound $l_b$ ($l_b \leq x \leq u_b$) as well as a unit cost $c$.
The total cost is of course given by summing $cx$ over all edges in $G$.

Suppose that the {\em bipartite} honeycomb lattice 
with periodic boundary conditions contains $2N$ sites which we 
divide into two sublattices of $N$ A-sites and $N$ B-sites. 
We can construct a graph $G$ as follows. In addition to all sites and bonds 
of the honeycomb lattice, this graph contains two extra
sites, denoted as $s$ (the source) and $t$ (the sink), 
and extra $2N$ bonds (the leads). All bonds of the honeycomb
lattice are directed from A-sites to B-sites with
$l_b=0$, $u_b=1$, and $c$ being the corresponding
random bond weight, while the remaining $N$ in-leads are
directed from $s$ to A-sites and $N$ out-leads from B-sites 
to $t$ with $l_u=1$, $u_b=2$, and $c=0$ for all
$2N$ leads. Therefore, the ground state energy of a loop 
configuration with or without defects is equivalent to
the minimum-cost flow if loops and strings are identified
with bonds on the honeycomb lattice that have flow (note that 
the flow value on these bonds must be either zero or unity).  
Simple inspection shows that this identification can indeed  
be made with the above choice of bounds if the amount of
flow sustained between $s$ and $t$, with flow conservation
on all other nodes, is between $N$ and $2N$ units. 

\begin{figure}[tbhp]
  \begin{center}
    \leavevmode
    \epsfxsize=7.5cm
    \epsffile{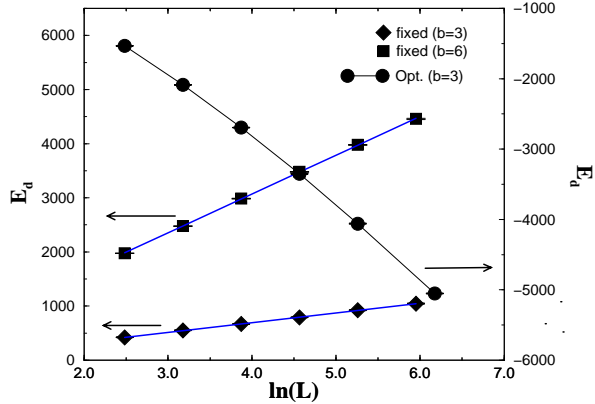}
  \end{center}
\caption{ 
Energetics of a dislocation pair. Diamond and square 
symbols denote the defect energy $E_d$ for a pair 
of {\it fixed} dislocations with the Burgers charges 
$\pm 3$ and $\pm 6$ respectively. Solid lines are 
linear fits. Data denoted by circles are the defect 
energy $E_d$ for a pair of {\it optimized} 
dislocations with the Burgers charges $\pm 3$.  
}
\label{fig_2}
\end{figure}

Given the amount of flow sustained, the minimum-cost-flow
algorithm establishes the flow pattern of the minimum 
cost. 
Various interesting physical situations can be 
simulated by simple variations.  
$2N$ units of flow, for example, lead to the ground state 
of fully-packed loops (no dislocations). $2N-1$ units of
flow, on the other hand, give the ground state with one 
dislocation pair without {\em a priori} fixing the pair  
location. Keeping $2N-1$ units of flow while 
changing $u_b$ of a particular in-lead and out-lead 
from $2$ to $1$ simulates a fixed pair of
dislocations with the Burgers charges $\pm 3$.  
If using $2N-2$ units of flow instead and changing $u_b$ of a particular
in-lead and out-lead from $2$ to $0$, we obtain 
a pair of dislocations (holes) with the Burgers charges $\pm 6$ 
at fixed locations. Clearly, dislocations of any desired 
density can be achieved by suitably varying the flow 
between $N$ and $2N$ units. Moreover, introducing 
another extra link from $t$ back to $s$ 
with a negative unit cost $-E_c$ allows us to determine 
the optimal amount of flow sustained (thus  
the optimal dislocation density) with $E_c$ being the 
core energy. This last simple variation results in the  
min-cost-circulation problem in network flow optimization.

\paragraph{Numerical results}  

For a given disorder realization, two ground-state energies,
$E_1$ and $E_0$, were obtained respectively for cases with
and without dislocations. The defect energy $E_d \equiv E_1-E_0$ 
was then determined. Various $L\times L$ sample sizes with $L=12,24,
48,96,192$, and $384$ ($480$ for optimized defects) were simulated 
with at least $10^4$ disorder averages for each size. 

We first describe our results for a single dislocation pair 
where the core energy $E_c$ is set to zero. 
The elastic constant of an elastic medium can be 
measured in various ways by observing its response to 
perturbations. Here we perturb the system 
with a fixed dislocation pair (large-scale topological 
excitation). The defect energy $E_d$ in this case is 
the elastic energy cost $E_{ela}$ which according to 
the elastic theory should scale as $E_{ela} \sim 
Kb^2/2\pi \ln(L)$. This is indeed consistent with our numerical
results shown in Fig.2, and the elastic constant $K$ is found  
to be $126(2)$ and $125(1)$ from dislocations pairs with the 
the Burgers charges $\pm 3$ and $\pm 6$ respectively. 
When the dislocation pair is allowed to be placed optimally, 
$E_d$ also contains the disorder energy gain $E_{dis}$ 
in addition to $E_{ela}$, i.e., $E_d = E_{ela}+ E_{dis}$. 
As shown clearly in Fig. 2, $E_{dis}$ dominates over $E_{ela}$ 
resulting in the negative $E_d$, and moreover, $E_{dis}$ 
drops faster than $\ln(L)$. Detailed analysis showed that 
the numerical results are consistent with the theoretical 
prediction $E_{dis} \sim -\ln^{3/2}(L)$\cite{dislocation1}, 
a result independent of the disorder strength $w$. Therefore
the elastic phase of large systems is unstable to 
dislocation pairs. With no restrictions on their number, 
dislocations will proliferate thereby driving the elastic 
constant $K$ to zero.      
 
\begin{figure}
  \begin{center}
    \leavevmode
    \epsfxsize=8cm
    \epsffile{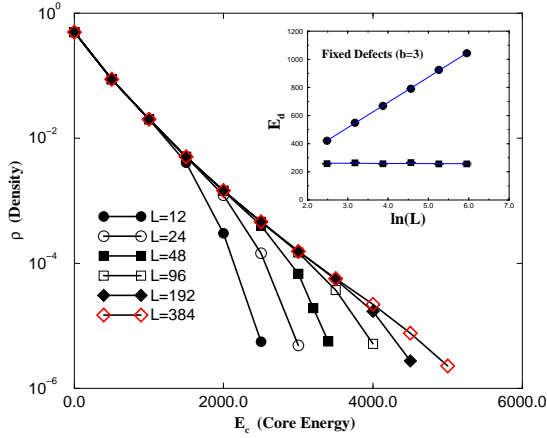}
  \end{center}
\caption{  
Optimal dislocation density $\rho$ as a function of 
the core energy $E_c$. Shown in the inset are the 
elastic-energy costs ($E_d = E_{ela}$) to a pair of fixed dislocations
with $E_c=0$ and the Burgers charges $\pm 3$ injected into a state 
with already optimal number of pre-existing dislocations (square) 
as well as into a state with no pre-existing dislocations
(circle). Data denoted by circles are the same 
as those denoted by diamonds in Fig. 2 and are 
shown here for comparison.}  
\label{fig_3}
\end{figure} 

We now discuss our results on multiple dislocations which 
are summarized in Fig. 3. It is indeed clear from 
the inset that the elastic energy cost $E_{ela}$ for introducing
a pair of fixed dislocations into the state where the 
number of dislocations is already optimal is independent 
of the separation of the fixed pair implying a zero elastic constant $K$.   
This is consistent with the result on a related model\cite{Multi}.  
The relation between the optimal dislocation density
$\rho$ and the core energy 
$E_c$ is, however, found to be  
\begin{equation} 
\rho \sim e^{-(E_c/E_0)^{\alpha}}  
\end{equation}  
with $\alpha = 0.74(3)$. This exponent remains elusive 
to us at the present.    
 
\paragraph {Conclusion and Outlook} 
In conclusion, we studied the energetics of dislocation pairs in 
a $2d$ random elastic medium by applying polynomial algorithms 
to $2d$ disordered FPL model and found the elastic phase is 
unstable against the proliferation of dislocations, and thus providing 
evidence against the formation of a Bragg glass in two dimensions. 

Further exploration of these disordered $2d$ lattice models with SOS 
representations will certainly help to address the fundamental issue 
of how non-random critical systems are affected by quenched
disorder since most of these non-random models are critical\cite{height}. 
For example, the non-random FPL model flows to the densely-packed 
loop (DPL) fixed point of the O(n) model upon the perturbation of 
holes\cite{DPL}. How this DPL fixed point get modified by 
bond randomness (if at all) can now be examined by using 
non-bipartite matching algorithms. 
Another exciting extension of these polynomial algorithms to compute  
the energetics of dislocation loops in $3d$ random elastic media 
is now also feasible. Compared with the past few years in which 
polynomial and {\it exact} algorithms have been productively explored,
it is fair to say that the next few years will see a rapid closing-in 
on a class of even NP-hard disordered systems which allow  
polynomial and {\it approximant} algorithms with near-optimal 
solutions of guaranteed bounds.   

We thank J. Kondev, C.L. Henley and A.A. Middleton for useful discussions. 
Part of this work is done in collaboration with D.S. Fisher which is also 
gratefully acknowledged.

\end{document}